\documentclass[twocolumn]{aastex631}

\usepackage{natbib}
\usepackage{multirow}
\usepackage{color}
\usepackage{bm}
\usepackage[normalem]{ulem}
\newcommand{\eg}{e.g., }
\newcommand{\ie}{i.e., }
\newcommand{\Msun}{M_{\odot}}

\newcommand{\Mej}{M_{\rm ej}}

\newcommand{\Ye}{Y_{\rm e}}
\def\gsim{\mathrel{\rlap{\lower 4pt \hbox{\hskip 1pt $\sim$}}\raise 1pt \hbox {$>$}}}
\def\lsim{\mathrel{\rlap{\lower 4pt \hbox{\hskip 1pt $\sim$}}\raise 1pt \hbox {$<$}}}

\shorttitle{Exploring Th features in Kilonova Spectra}
\shortauthors{N. Domoto et al.}
\begin{document}

\title{Thorium in Kilonova Spectra: Exploring the Heaviest Detectable Element}

\correspondingauthor{Nanae Domoto}
\email{n.domoto@astr.tohoku.ac.jp}

\author[0000-0002-7415-7954]{Nanae Domoto}
\affiliation{Astronomical Institute, Tohoku University, Aoba, Sendai 980-8578, Japan}

\author[0000-0002-4759-7794]{Shinya Wanajo}
\affiliation{Astronomical Institute, Tohoku University, Aoba, Sendai 980-8578, Japan}
\affiliation{Max-Planck-Institut f\"{u}r Gravitationsphysik (Albert-Einstein-Institut), Am M\"{u}hlenberg 1, D-14476 Potsdam-Golm, Germany}
\affiliation{Interdisciplinary Theoretical and Mathematical Sciences Program (iTHEMS), RIKEN, Wako, Saitama 351-0198, Japan}

\author[0000-0001-8253-6850]{Masaomi Tanaka}
\affiliation{Astronomical Institute, Tohoku University, Aoba, Sendai 980-8578, Japan}
\affiliation{Division for the Establishment of Frontier Sciences, Organization for Advanced Studies, Tohoku University, Sendai 980-8577, Japan}

\author[0000-0002-5302-073X]{Daiji Kato}
\affiliation{National Institute for Fusion Science, 322-6 Oroshi-cho, Toki 509-5292, Japan}
\affiliation{Interdisciplinary Graduate School of Engineering Sciences, Kyushu University, Kasuga, Fukuoka 816-8580, Japan}

\author[0000-0002-2502-3730]{Kenta Hotokezaka}
\affiliation{Research Center for the Early Universe, Graduate School of Science, University of Tokyo, Bunkyo, Tokyo 113-0033, Japan}
\affiliation{Kavli IPMU (WPI), UTIAS, The University of Tokyo, Kashiwa, Chiba 277-8583, Japan}



\begin{abstract}
Kilonova spectra provide us with the direct information of $r$-process nucleosynthesis in neutron star mergers.
In this paper, we study the signatures of elements beyond the third $r$-process peak expected to be produced in neutron-rich ejecta in the photospheric spectra of kilonova.
Ra II, Ac III, and Th III are our candidates because they have a small number of valence electrons and low-lying energy levels, which tend to result in strong absorption features.
We systematically calculate the strength of bound-bound transitions of these candidates by constructing the line list based on the available atomic database.
We find that Th III is the most promising species showing strong transitions at the NIR wavelengths.
By performing radiative transfer simulations, we find that Th III produces broad absorption features at $\sim$18000 {\AA} in the spectra when the mass ratio of actinides to lanthanides is larger than the solar $r$-process ratio and the mass fraction of lanthanides is $\lesssim 6\times10^{-4}$.
Our models demonstrate that the Th feature may be detectable if the bulk of the ejecta in the line-forming region is dominated by relatively light $r$-process elements with the mixture of a small fraction of very neutron-rich material.
Such conditions may be realized in the mergers of unequal-mass neutron stars or black hole-neutron star binaries.
To detect the Th absorption features, the observations from the space (such as JWST) or high-altitude sites are important as the wavelength region of the Th features is overlapped with that affected by the strong telluric absorption.
\end{abstract}


%\keywords{editorials, notices --- miscellaneous --- catalogs --- surveys}

\section{Introduction}
\label{sec:intro}
Coalescence of binary neutron stars (NSs) is a promising site of the rapid neutron capture nucleosynthesis \citep[$r$-process, e.g.,][]{LS1974, Eichler1989, Freiburghaus1999, Goriely2011a, Korobkin2012, Wanajo2014}.
Radioactive decay of freshly synthesized nuclei in the ejected neutron-rich material powers electromagnetic emission called a kilonova \citep{LiPaczynski1998, Metzger2010, Roberts2011}.
In 2017, associated with the detection of gravitational waves (GW) from an NS merger \citep[GW170817,][]{Abbott2017a}, an electromagnetic counterpart was identified \citep[AT2017gfo,][]{Abbott2017b}.
The observed properties of AT2017gfo at ultraviolet, optical, and near-infrared (NIR) wavelengths are consistent with the theoretical expectation of a kilonova \citep[e.g.,][]{Arcavi2017, Coulter2017, Evans2017, Pian2017, Smartt2017, Utsumi2017, Valenti2017}.
The electromagnetic counterpart has provided us with evidence that NS mergers are sites of $r$-process nucleosynthesis \citep[e.g.,][]{Kasen2017, Perego2017, Shibata2017, Tanaka2017, Kawaguchi2018, Rosswog2018}.

Recent advances in the study of kilonova spectra have enabled the direct identification of individual elements.
In the photospheric spectra at an early phase, one can identify the synthesized elements from the absorption features.
\citet{Watson2019} first reported that the absorption features around $\lambda\sim8000$ {\AA} in the photospheric spectra of AT2017gfo could be explained by Sr II ($Z=38$).
This has been further supported by independent analyses \citep{Domoto2021, Gillanders2022}, although the same features may be caused by He I (\citealp{Perego2022, Tarumi2023}, but see also \citealp{Sneppen2024}).
\citet{Domoto2022} reported the absorption features at $\lambda\sim13000$ and 14500 {\AA} in the spectra of AT2017gfo as to be caused by La III ($Z=57$) and Ce III ($Z=58$), respectively.
The identification of Ce III has been further supported by another approach using the stellar spectra that show strong Ce III absorption \citep{Tanaka2023, Domoto2023}.
Furthermore, \citet{Sneppen2023} interpreted the marginal absorption features at $\lambda\sim7600$ {\AA} caused by Y II ($Z=39$), although \citet{Pognan2023} suggested the same features due to Rb I ($Z=37$).

In the nebula spectra at a late phase, one can also identify the elements from emission features.
\citet{Hotokezaka2023} studied the spectra of AT2017gfo at $>7$ days after the merger, and reported the emission features at $\lambda\sim 2.1$ {\textmu}m as to be caused by Te III ($Z=52$).
In GW170817, the detection of nebula emission at 4.5 {\textmu}m and the upper limit at 3.6 {\textmu}m by the Spitzer space telescope at very late phases were reported \citep{Villar2018, Kasliwal2022}, which suggests the distinctive spectral shape.
It has been suggested that these features may be explained by Se ($Z=34$) or W ($Z=74$) \citep{Hotokezaka2022}, although conclusive spectroscopic identification has not been made.

Despite the identification of these elements in the spectra of AT2017gfo, the overall abundance pattern in this NS merger event is not yet clear.
It is expected that so-called third $r$-process peak elements, \eg Pt and Au ($Z=78$, 79), have been abundantly synthesized by this event \citep[e.g.,][]{Goriely2011a, Korobkin2012, Wanajo2014}.
In fact, \citet{Gillanders2021} searched the signatures of Pt and Au in the spectra of AT2017gfo, but concluded that these elements do not produce prominent line features.
There is currently no established way to identify these elements with spectroscopic observations.

The very neutron-rich ejecta of NS merger is expected to produce actinides \citep[e.g.,][]{Fujibayashi2023, Kullmann2023}.
It has been proposed that one can trace such heavy nuclei by the late-time bolometric light curve, as the radioactive heating is often dominated by a few isotopes with the half-lives of 10--100 days \citep{Zhu2018, Wu2019}.
However, such constraint was difficult for AT2017gfo due to the limited observations in the late phase.
Also, \citet{Domoto2022} briefly discussed the spectral feature caused by an actinide Th ($Z=90$).
They found that Th produces the absorption feature at $\lambda\sim18000$ {\AA}, although the feature is not as prominent as those of lanthanides.
However, they used only one abundance model dominated by relatively light $r$-process elements and did not investigate the conditions that give rise to the Th features in detail.
Furthermore, they did not investigate the features of heavy elements around actinides other than Th, such as Ra and Ac.
Since the detection of such elements provides us with unambiguous evidence that heavy $r$-process nuclei are synthesized in NS mergers, it is important to clarify the detectability of these features.

In this paper, we explore the detectability of absorption features caused by the elements beyond the third $r$-process peak in photospheric spectra of kilonvoae.
In particular, motivated by the finding in \citet{Domoto2022}, we study the absorption features caused by Th III at $\lambda\sim18000$ {\AA} in the spectra in detail.
In Section \ref{sec:candidate}, we first systematically calculate the strength of bound-bound transitions of candidate species to show that Th III is the most plausible element among candidates as the source of absorption features.
Then, in Section \ref{sec:spec}, we perform radiative transfer simulations of NS merger ejecta and investigate the conditions in which we can find the absorption features of Th III.
The implication and caveat of our results are discussed in Section \ref{sec:discussion}.
Finally, we give our conclusions in Section \ref{sec:conclusion}.

\section{Th III and other candidate species}
\label{sec:candidate}
%===========================
% Figure
%===========================
\begin{figure}[th]
  \begin{center}
    \includegraphics[width=\linewidth]{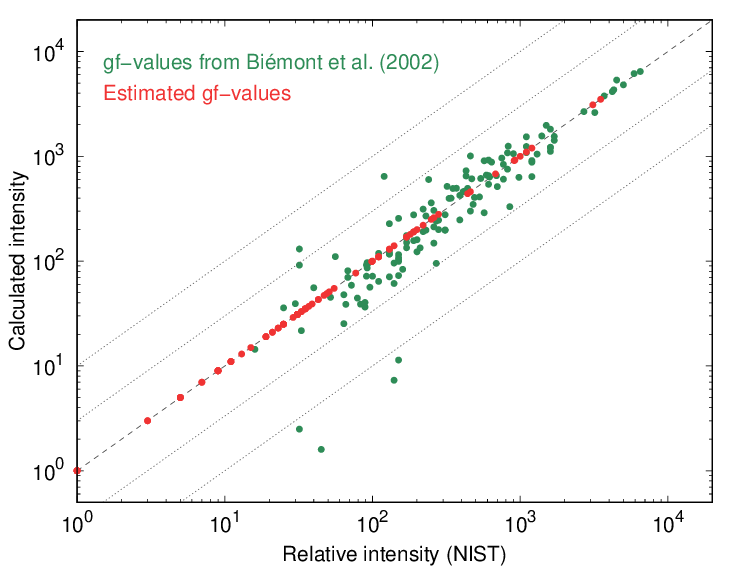}
\caption{
  \label{fig:gf}
  Comparison of intensities (green circles) for Th III lines between those calculated with $gf$-values from \citet{Biemont2002} and those measured by experiments \citep{Engleman2003, nist}. 
  Gray dashed and dotted lines correspond to the perfect agreement and the deviations by a factor of 3 and 10, respectively.
  Red circles indicate the lines whose $gf$-values are estimated from the measured intensities.
}
\end{center}
\end{figure}
%===========================
\begin{figure*}[th]
  \begin{center}
    \begin{tabular}{cc}
    \includegraphics[width=0.48\linewidth]{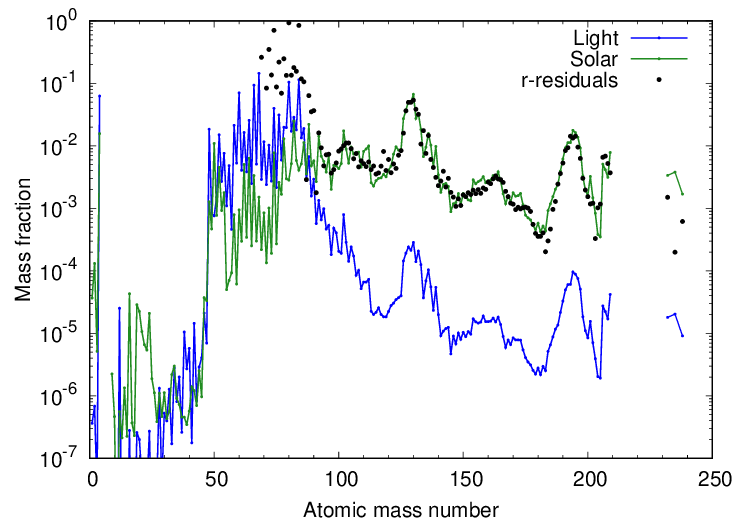} &
    \includegraphics[width=0.48\linewidth]{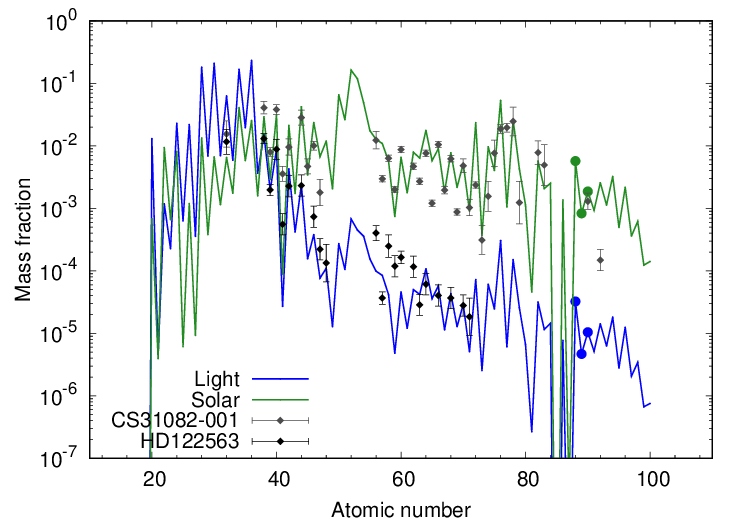}
    \end{tabular}
\caption{
  \label{fig:abun}
  Left: final abundances (at 1 yr; all trans-Pb nuclei except for Th and U are assumed to have decayed) of the L (blue) and S (green) models as a function of mass number.
  Black circles show the $r$-process residual pattern \citep{Prantzos2020}, which are scaled to match those for the S model at $A = 138$.
  Right: abundances at $t=1.5$ days for each model as a function of atomic number.
  Abundances of an $r$-process-deficient star HD 122563 (black diamonds, \citealp{Honda2006}; Ge from \citealp{Cowan2005}; Cd and Lu from \citealp{Roederer2012}) and an r-process-enhanced star CS 31082-001 \citep[gray diamonds; ][]{star} are also shown for comparison purposes.
  The abundances of HD 122563 and CS 31082-001 are scaled to those for the L and S models at $Z=40$, respectively. 
  The elements relevant to this study are indicated by large circles in each model (see also Table \ref{tab:abun}).
}
\end{center}
\end{figure*}
%===========================

We first consider the heaviest candidate species that may produce absorption features in kilonova photospheric spectra.
It has been shown that the elements on the left side of the periodic table, such as Ca, Sr, La, and Ce tend to produce absorption features in the spectra \citep{Domoto2022}.
This is explained by atomic properties: 
a small number of valence electrons for which transition probabilities ($gf$-values) tend to be larger, and low-lying energy levels for which the electron population tends to be higher (see Section \ref{sec:strength}).
Among the elements beyond the third $r$-process peak, Ra II ($Z=88$), Ac III ($Z=89$), and Th III ($Z=90$) are plausible candidates to produce absorption features, because their atomic structures are expected to be analogous to those of Ca II/Sr II, La III, and Ce III, respectively.
Note that Ra and Ac are radioactive elements, but their isotopes with relatively long half-lives (an order of days or longer) can exist in NS merger ejecta at a few days after the merger.
In the following subsections, however, we will show that Th III is the only plausible candidate for the absorption features.

\subsection{Line list}
\label{sec:linelist}
%===========================
% Table 
%===========================
\begin{deluxetable*}{lcccccc}[th]
\tablewidth{0pt}
\tablecaption{Mass fractions of selected elements.}
\label{tab:abun}
\tablehead{
 Model &  &  $X$(Ra) & $X$(Ac) & $X$(Th) & $X$(Ce) & $X$(lan)$^a$ 
}
\startdata
Light (L)  & Final       &                -               &                 -              & 1.8$\times10^{-5}$ &  3.1$\times10^{-5}$ & 4.6$\times10^{-4}$  \\ %final
               &1.5 days  &  3.2$\times10^{-5}$ & 4.7$\times10^{-6}$ & 1.0$\times10^{-5}$ &  4.2$\times10^{-5}$ & 5.8$\times10^{-4}$  \\ \hline   %t=1.5d
Solar (S) & Final       &                -               &                 -              & 3.4$\times10^{-3}$ &  5.9$\times10^{-3}$ & 8.5$\times10^{-2}$  \\ %final
               &1.5 days  &  5.7$\times10^{-3}$ & 8.3$\times10^{-4}$ & 1.9$\times10^{-3}$ &  5.7$\times10^{-3}$ & 9.0$\times10^{-2}$     %t=1.5d
\enddata
\tablecomments{
The top and bottom rows in each model show the final abundances and those at $t=1.5$ days, respectively.\\
$^a$ Sum of mass fractions for lanthanides ($Z=57$--71).
}
\end{deluxetable*}
%===========================

To evaluate the strength of bound-bound transitions, relevant atomic data, \ie a set of transition wavelengths, energy levels of transitions, and transition probabilities ($gf$-values) are necessary.
In this paper, we use the ``hybrid'' line list from \citet{Domoto2022, Domoto2023} as a baseline.
This line list consists of the theoretically calculated data for the elements of $Z=30$--88 \citep{Tanaka2020} as well as experimentally calibrated data for the elements of $Z=20$--29 \citep{Piskunov1995, Kupka1999, Ryabchikova2015} and for several important ions for the absorption features in kilonovae, Sr II, Y I, Y II, Zr I, Zr II, Ba II, La III, and Ce III.

Although the hybrid line list contains the data for Ra II from the theoretical calculations, the data were not experimentally calibrated as it was outside the scope of the study in \citet{Domoto2022, Domoto2023}.
Fortunately, some experimental data of the transitions for Ra II are available in the NIST Atomic Spectral Database \citep[ASD,][]{nist}.
Some of them have the $gf$-values from theoretical calculations \citep{RaII} with high accuracy.
Thus, we replace the Ra II data in the hybrid line list with the data in the NIST ASD for which the $gf$-values are available.
The experimental transition data for Ac III are also available in the NIST ASD, with the $gf$-values from theoretical calculations \citep{AcIII1, AcIII2} with high accuracy \citep{Kramida2022}.
Thus, we also add the available data for Ac III from the NIST ASD to our line list.

For Th III, experimental data for the transition wavelengths and energy levels are available for the optical to NIR region \citep{Engleman2003}.
However, while the $gf$-values of optical lines are obtained by using semi-empirical calculations \citep{Biemont2002}, those of NIR lines are not available.
\citet{Domoto2022} estimated the $gf$-values of the NIR lines using the measured relative intensities $I$ by 
\begin{eqnarray}
	I &=& b\ g_u A\ e^{-\frac{E_u}{kT}} \nonumber \\
	  &=& b \frac{8\pi e^2}{m_e c\lambda^2}g_l f_{\rm osc} e^{-\frac{E_u}{kT}},
	\label{eq:I}
\end{eqnarray}
where $A$, $\lambda$, and $f_{\rm osc}$ are the Einstein's A coefficient, the transition wavelength, and the oscillator strength, $g$ and $E$ are the statistical weight and the energy level of the lower ($l$) and upper ($u$) levels for a transition, respectively, and $b$ is a constant depending on the ion species.
This equation is valid since the electron population in the experiments follows the Boltzmann distribution with atomic excitation temperature \citep{Thatlas}.
As the relative intensities of the measured lines are listed in the NIST ASD in a consistent way through the whole wavelength range, one can compare the calculated and measured intensities for the lines over a wide range of wavelengths.

We update the $gf$-values of the NIR lines for Th III by following \citet{Domoto2022}, but in a more quantitative way.
To calculate the intensities of the optical lines for Th III with known $gf$-values \citep{Biemont2002} using Equation (\ref{eq:I}), 
\citet{Domoto2022} adopted the temperature $T=5000$ K by a visual inspection, at which the measured and calculated intensities of the optical lines agree with each other.
Instead, here we use the least squares method to find the temperature at which the relationship of the measured and calculated intensities are closest to be linear.
As a result, the temperature becomes $T=6000$ K, which we adopt in this study.
A comparison between the measured intensities and calculated intensities using the known $gf$-values is shown in Figure \ref{fig:gf}.
Then, we estimate the $gf$-values of the NIR lines for Th III by using this temperature of $T=6000$ K.
Consequently, the estimated values of the $gf$-values are systematically smaller than those in \citet{Domoto2022} by up to a factor of $\sim$3.
We summarize the estimated $gf$-values in Table \ref{tab:gf}.\footnote{As all the $g_l$-values ($=2J_l+1$, where $J_l$ is the total angular momentum of the lower level) are known for the NIR lines of Th III (Table \ref{tab:gf}, see the NIST ASD for the complete information), the procedure in fact gives the $f$-values not $gf$-values for all the NIR lines.}

Note that there is a scatter between measured and calculated intensities by roughly a factor of 3, especially for the weaker lines.
This can be due to the uncertainty in the used $gf$-values, which is expected to be 5--10\% or more \citep{Biemont2002}.
However, it is difficult to assess the uncertainty with lack of experimental evaluations.
Also, the presence of the scatter might also indicate that the electron population in the experiments does not completely follow the Boltzmann distribution.
These may cause the uncertainty in the temperature estimates and resulting $gf$-values of the NIR lines (see also Section \ref{sec:discussion}).

We mention that there are some theoretical atomic data of actinides available \citep{Silva2022, Floers2023, Fontes2023}.
Such data are constructed performing atomic calculations to estimate the opacity and its effect on kilonovae.
Such theoretical data, however, are not calibrated with experiments, and thus, they cannot be readily used to discuss spectral features.
Nevertheless, those studies have shown that actinides tend to have many low-lying energy levels than lanthanides, which is consistent with the experimental results for Th III (see the next section).

%===========================
% Figure
%===========================
\begin{figure}[th]
  \begin{center}
    \includegraphics[width=\linewidth]{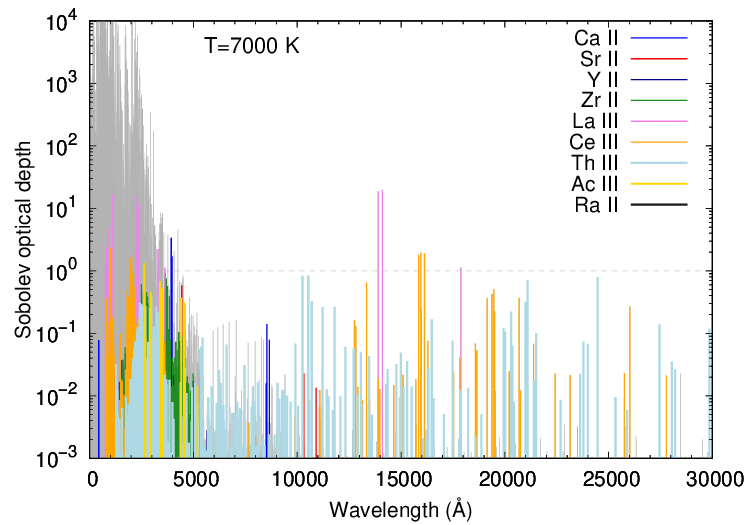} \\
    \includegraphics[width=\linewidth]{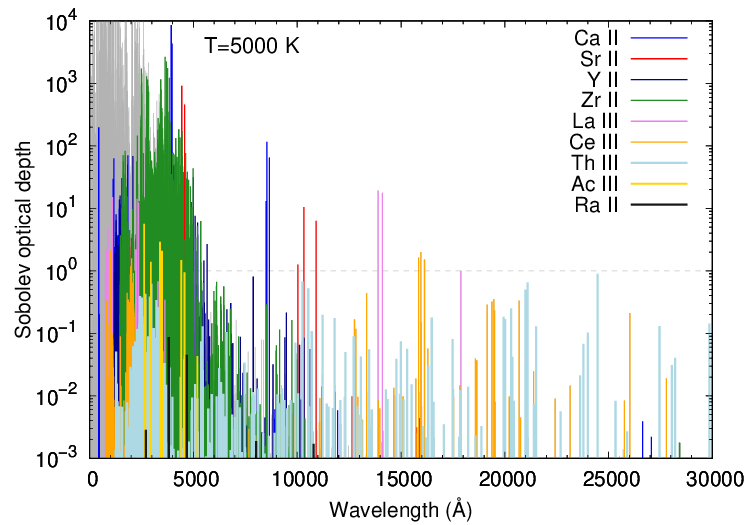} \\
    \includegraphics[width=\linewidth]{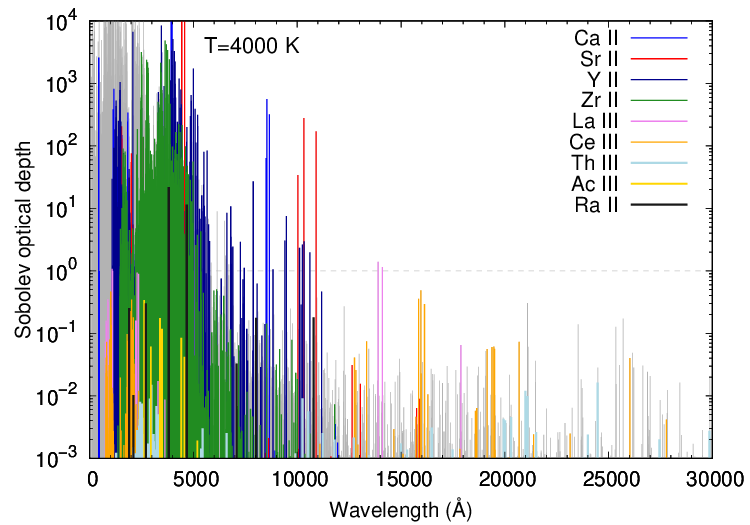}  \\
\caption{
  \label{fig:tau}
  Sobolev optical depth of bound-bound transitions for the L model.
  The ions with large contributions and of interest are shown with colors.
  The top to bottom panels show the results with the temperature of $T=7000$, 5000, and 4000 K, respectively, and the density of $\rho=10^{-14}$ g~cm$^{-3}$ at $t=1.5$ days.
}
\end{center}
\end{figure}
%===========================

\subsection{Strength of lines for Ra, Ac, and Th}
\label{sec:strength}
To investigate the important species for absorption features in kilonova spectra among the heaviest elements, we systematically calculate the strength of bound-bound transitions for a given set of density, temperature, and element abundances.
The strength of a line is approximated by the Sobolev optical depth \citep{Sobolev1960} for each bound-bound transition,
\begin{eqnarray}
	\tau_{\rm sob} &=& \frac{\pi e^2}{m_e c} n_{i, j, k} t \lambda f_{\rm osc}
	\label{eq:tau}
\end{eqnarray}
in homologously expanding ejecta.
The Sobolev approximation is valid for the matter with a high expansion velocity and a large radial velocity gradient.
Here, $n_{i, j, k}$ is the number density of the ion at the lower level of a transition ($i$-th element, $j$-th ionization stage, and $k$-th excited state).
As in previous work of kilonovae \citep[e.g.,][]{Kasen2013, TH2013}, we assume LTE;
we solve the Saha equation to obtain ionization states, and assume the Boltzmann distribution for the population of excited levels \citep[see][for non-LTE effects]{Hotokezaka2021, Pognan2022a}.

For the abundances in the ejected matter from an NS merger, we use the same model as in \citet{Domoto2022}, which is based on a multicomponent free-expansion (mFE) model of \citet{Wanajo2018}.
The model is constructed as an ensemble of parametrized outflows for each constant velocity, initial entropy, and initial electron fraction ($\Ye$).
Here, we use the Light (L) model adopted in \citet{Domoto2022} as our fiducial model, which approximately fits the $r$-process residuals \citep{Prantzos2020} for $A=66$--69 and 3\% of those for $A=100$--205, where $A$ is atomic mass number (the blue curves in Figure \ref{fig:abun}).
This is motivated by the fact that the L model can reasonably explain the observed features of heavy elements such as La III and Ce III in AT2017gfo \citep{Domoto2022}.
We refer the readers to Section \ref{sec:spec} for the discussion using a different abundance distribution.
It should be noted that we do not intend this abundance pattern to represent the whole ejecta but those in the line-forming regions.
We use the abundances at $t=1.5$ days for $Z=20$--100 in our calculations.
The mass fractions of elements relevant to this study are summarized in Table \ref{tab:abun}.

\begin{figure}[th]
  \begin{center}
    \includegraphics[width=\linewidth]{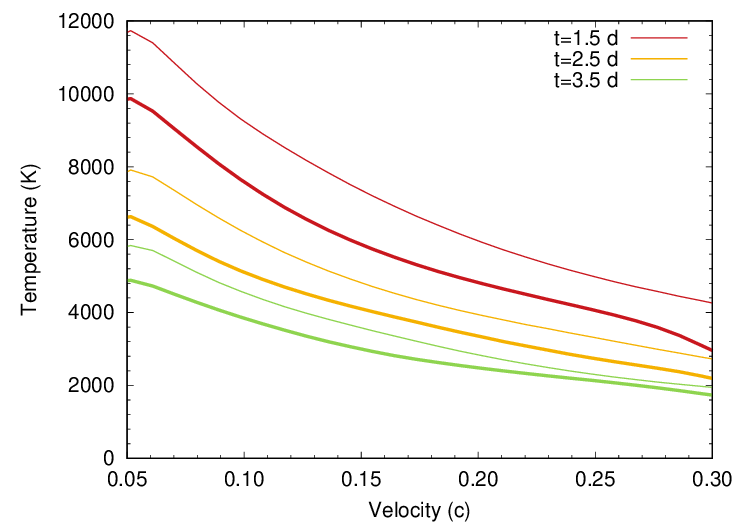}
\caption{
  \label{fig:temp}
  Temperature structure of the ejecta at $t=1.5$, 2.5, and 3.5 days after the merger for the L (thick lines) and S (thin lines) models.
}
\end{center}
\end{figure}

The Sobolev optical depths of bound-bound transitions for the L model at $t=1.5$ days are shown in Figure \ref{fig:tau}.
We evaluate the Sobolev optical depth for the density of $\rho=10^{-14}$ g~cm$^{-3}$ and the temperature of $T=7000$ and 5000 K (the top two panels).
These are typical values in the line-forming region for the L model at 1--2 days after the merger (see Figure \ref{fig:temp}).
As already shown in \citet{Domoto2022}, Ca II, Sr II, Y II, Zr II, La III, and Ce III show strong transitions.

%===========================
% Figure 
%===========================
\begin{figure}[th]
  \begin{center}
    \includegraphics[width=\linewidth]{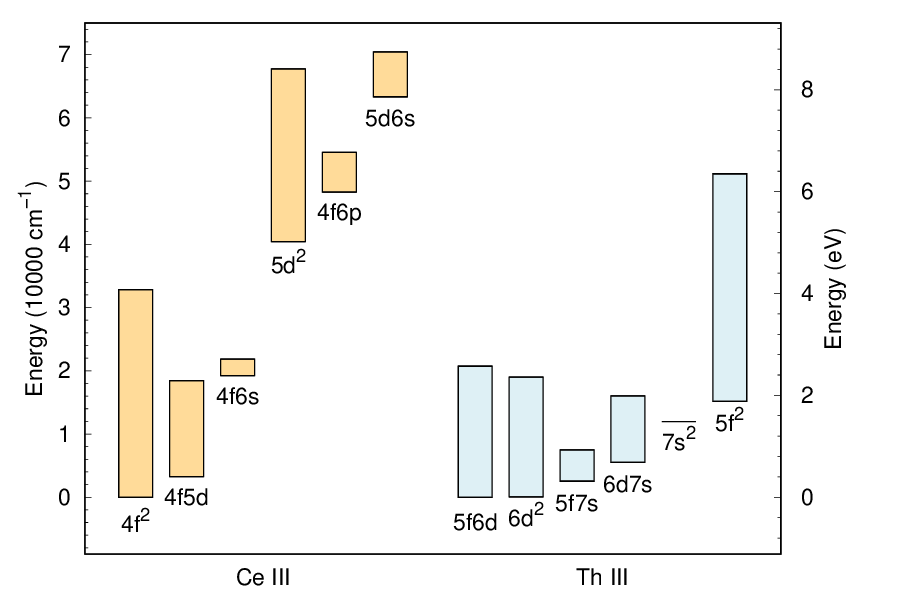}
\caption{
  \label{fig:Thlevel}
  Energy distributions of the six low-lying configurations for Ce III (left) and Th III (right).
  The bottom and top edges in each box correspond to the lowest and highest energies in each configuration, respectively.
  The first two configurations for Ce III are involved in the NIR region, while all the six configurations for Th III are involved in the NIR region.
}
\end{center}
\end{figure}
%===========================
\begin{figure}[th]
  \begin{center}
    \includegraphics[width=\linewidth]{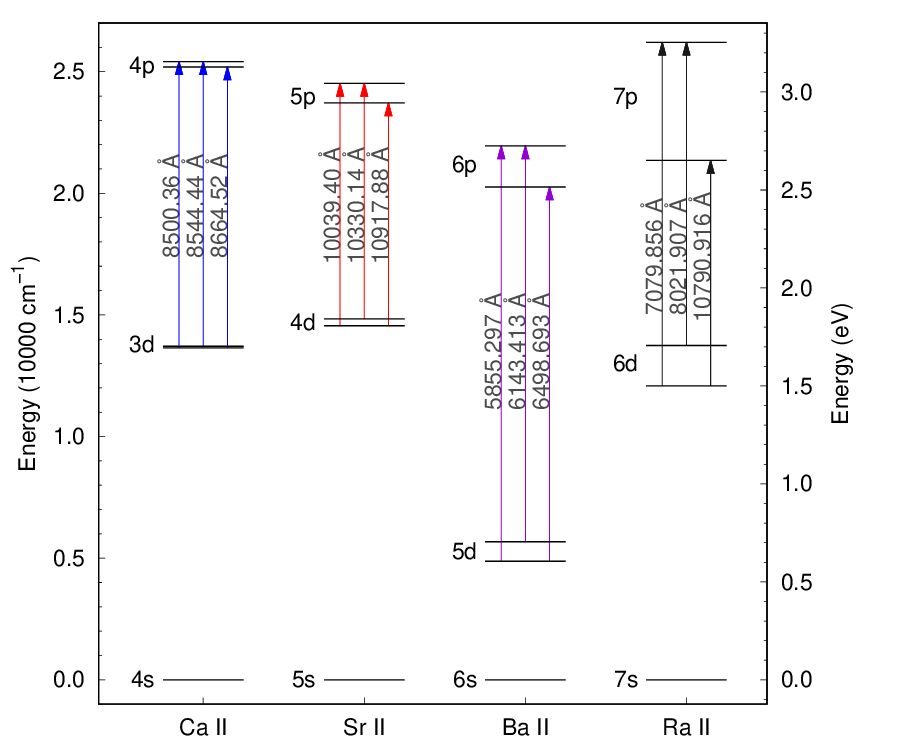}
\caption{
  \label{fig:triplet}
  Energy diagrams for Ca II, Sr II, Ba II, and Ra II.
  Each arrow shows the triplet transition with the value of the transition wavelength (in vacuum).
  The energy terms for these triplet lines are $^2$D$_{3/2}$--$^2$P$^o_{3/2}$, $^2$D$_{5/2}$--$^2$P$^o_{3/2}$, and $^2$D$_{3/2}$--$^2$P$^o_{1/2}$ from the shorter to longer wavelengths.
}
\end{center}
\end{figure}
%===========================
\begin{figure}[th]
  \begin{center}
    \includegraphics[width=\linewidth]{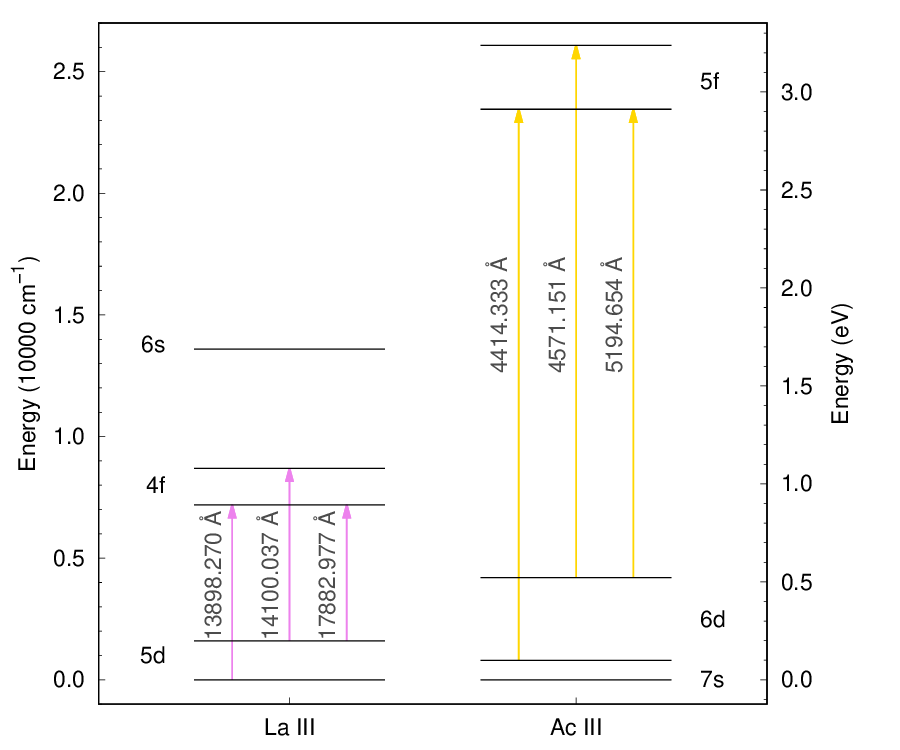}
\caption{
  \label{fig:Aclevel}
  Energy diagrams for La III and Ac III.
  Each arrow shows the transition with the value of the transition wavelength (in vacuum).
  The energy terms for these transitions are $^2$D$_{3/2}$--$^2$F$^o_{5/2}$, $^2$D$_{5/2}$--$^2$F$^o_{7/2}$, and $^2$D$_{5/2}$--$^2$F$^o_{5/2}$ from the shorter to longer wavelengths.
}
\end{center}
\end{figure}
%===========================

We find that the strength of the Th III lines (light-blue) at the NIR wavelengths are comparable to that of the Ce III lines (orange).
This is consistent with the finding in \citet{Domoto2022}, although the updated $gf$-values are adopted here.
As briefly discussed in \citet{Domoto2022}, many Th III lines show similar Sobolev optical depths and none of them are outstanding.
This is due to the fact that Th III has denser low-lying energy levels involved in the 5$f$-shell compared to those of Ce III involved in the 4$f$-shell (Figure \ref{fig:Thlevel}).
The Boltzmann factors at each level do not vary significantly among such low-lying levels, which makes the Sobolev optical depths of these lines similar.
The variations in the Sobolev optical depths mainly reflect the variations in $gf$-values, but in particular for the NIR range, the low-lying energy levels (as well as relatively high $gf$-values) enable the Th III NIR lines involving such levels (higher level population) to show relatively large Sobolev optical depths.
This is especially the case for the lines around $\lambda\sim20000$ {\AA}, which will be further investigated in Section \ref{sec:spec}.

For Ra II, it has the triplet lines similar to those of Ca II, Sr II, and Ba II, since their atomic structures having one electron in the $s$ orbit are analogous to each other (Figure \ref{fig:triplet}).
Nevertheless, the Ra II triplet lines do not show large Sobolev optical depths due to much smaller mass fraction of Ra than those of Ca or Sr by a factor of 10$^3$ (Figure \ref{fig:abun}).
From a nucleosynthesis point of view, it is not conceivable that Ra is more abundant than lanthanides or actinides by orders of magnitude.
In the case of the solar ratio (see Table \ref{tab:abun} and Figure \ref{fig:abun}), the mass fraction of Ra relative to individual lanthanides or Th are similar within a factor of a few at $t=1.5$ days.
Lanthanides have many lower energy levels below $\sim$2 eV, leading to larger Sobolev optical depths than those from Ra with similar mass fractions.
Thus, even if the mass fraction of Ra is somewhat larger than that in the L model assumed here, it would not be expected that the strength of the Ra II triplet lines can compete with those of other lines for lanthanides or Th III.
To fully verify this, however, we need experimentally calibrated data for more lanthanides at $\lambda\lesssim 10000$ {\AA}.

%===========================
% Figure 
%===========================
\begin{figure}[th]
  \begin{center}
    \includegraphics[width=\linewidth]{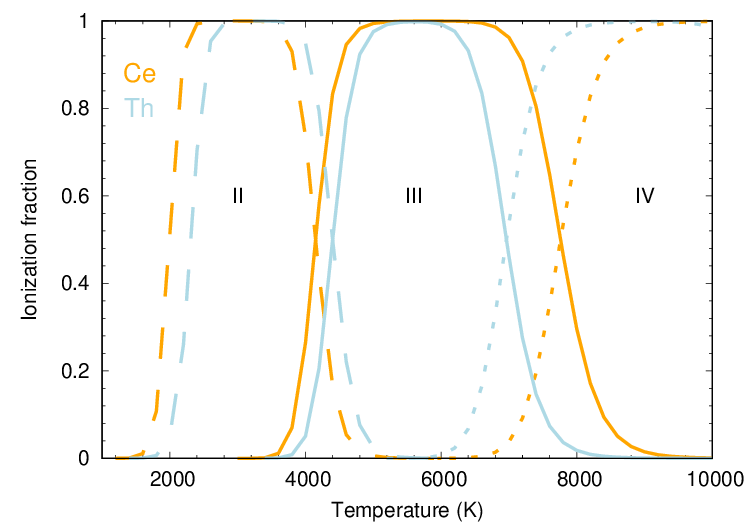}
\caption{
  \label{fig:ion}
  Ionization fractions of Ce (orange) and Th (light-blue) under LTE, assuming the density of $10^{-14}$ g~cm$^{-3}$.
  The dashed, solid, and dotted lines show the fractions of II, III, and IV, respectively.
}
\end{center}
\end{figure}
%===========================
%===========================

For Ac III, the available data reside in the forest of lines at the UV and optical wavelengths (yellow lines in Figure \ref{fig:tau}).
The strength of the Ac III lines are at most comparable to those of the other optical lines of heavy elements for higher temperature.
One might expect that the atomic structure of Ac III is analogous to that of La III that shows strong transitions at the NIR wavelengths.
However, it is not the case according to experimentally constructed atomic data \citep{Meggers1957, nist}:
while the ground-state electron configuration of La III is 5$d$, that of Ac III is 7$s$ (Figure \ref{fig:Aclevel}).
In the case of La III, the transitions from the ground configurations (5$d$) to the first excited configuration (4$f$) are very strong at the NIR wavelengths.
On the other hand, the energy differences between the 6$d$ and 5$f$ orbits for Ac III are large, which lead to the transition wavelengths being at the optical region.
Thus, Ac III does not have (allowed) lines at the NIR wavelengths involved in low-lying energy levels.

In summary, among the elements beyond the third $r$-process peak ($Z=88$--90), Th III is the most promising species that exhibits strong transitions in the NIR wavelengths under the relevant physical conditions of NS merger ejecta.
This is because Th III has many NIR lines involved in the dense low-lying energy levels.
In general, the number of lines is relatively small in the NIR region compared to that in the optical region.
Thus, individual NIR lines tend to be separated, resulting in recognizable features in the spectra.
Hereafter, we focus on Th III and explore the absorption features caused by Th III at the NIR region.

\subsection{Temperature dependence of Th III lines}
\label{sec:temp}
%===========================
% Figure 
%===========================
\begin{figure*}[th]
  \begin{center}
    \begin{tabular}{cc}
    \includegraphics[width=0.48\linewidth]{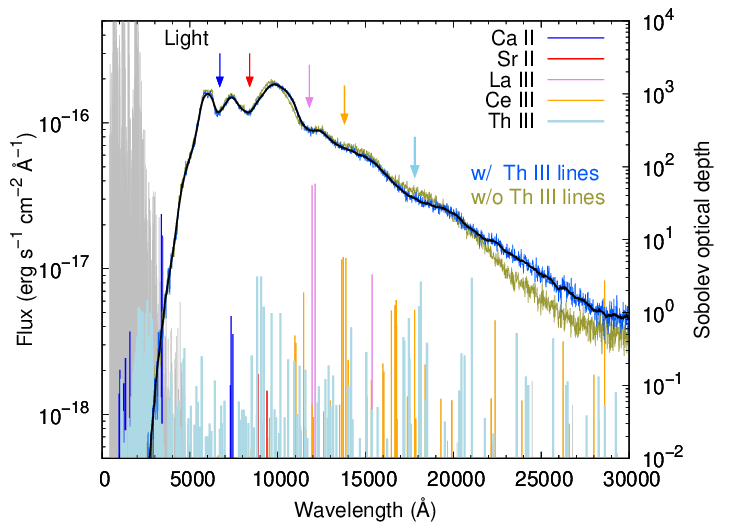} &
    \includegraphics[width=0.48\linewidth]{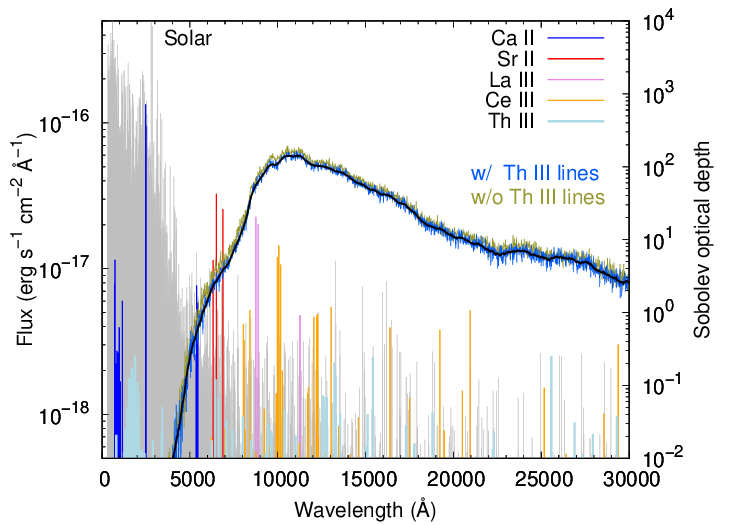}
    \end{tabular}
\caption{
  \label{fig:tau-spec}
  Left:
  synthetic spectrum (blue and black curves as original and smoothed results, respectively) and Sobolev optical depth of each transition (vertical lines) at $t=1.5$ days for the L model.
  The flux is presented by assuming the source at 40 Mpc as that for GW170817/AT2017gfo.
  We plot the Sobolev optical depths of spectroscopically accurate lines in the ejecta at $v=0.13\ c$.
  The positions of lines are blueshifted according to $v=0.13\ c$.
  The temperature in the ejecta at $v=0.13\ c$ is $T\sim7000$ K. 
  The synthetic spectrum without Th III lines is also shown for comparison purposes (dark-yellow curve).
  The absorption feature produced by each element is indicated by an arrow with the same color in the legend.
  Note that the Ca II and Sr II features (indicated by blue and red arrows) show too large Doppler shifts to be consistent with the positions of their lines.
  Right:
  same as the left panel but for the S model.
  We plot the Sobolev optical depths of the lines in the ejecta at $v=0.37\ c$, whose wavelengths are not necessarily accurate.
  The positions of lines are blueshifted according to $v=0.37\ c$.
  The temperature in the ejecta at $v=0.37\ c$ is $T\sim3700$ K. 
}
\end{center}
\end{figure*}
%===========================
\begin{figure*}[th]
  \begin{center}
    \begin{tabular}{cc}
    \includegraphics[width=0.48\linewidth]{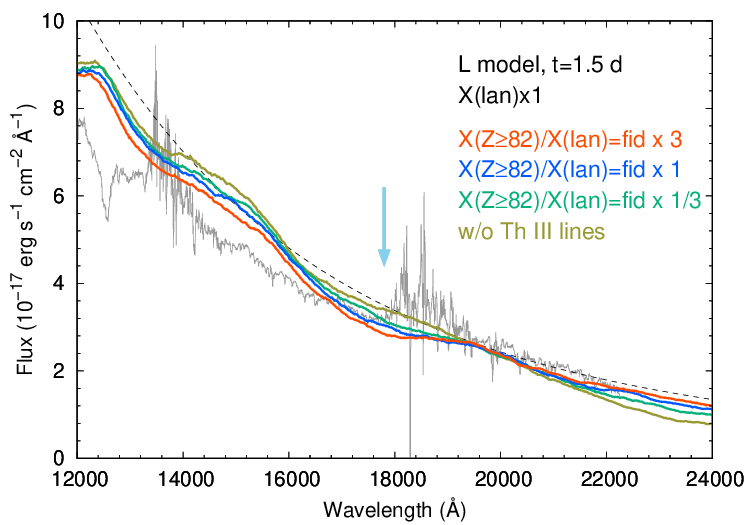} &
    \includegraphics[width=0.48\linewidth]{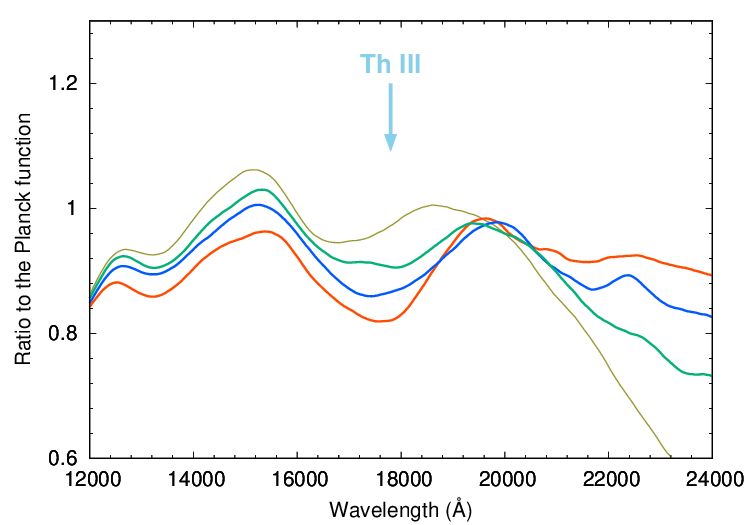} \\
    \includegraphics[width=0.48\linewidth]{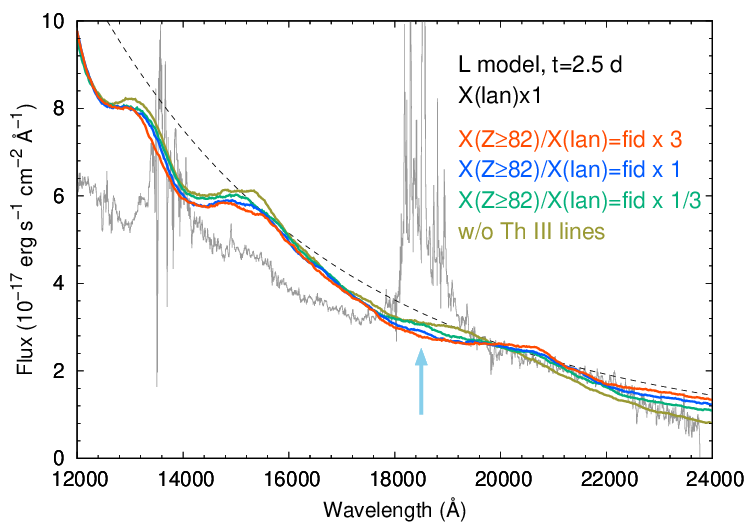} &
    \includegraphics[width=0.48\linewidth]{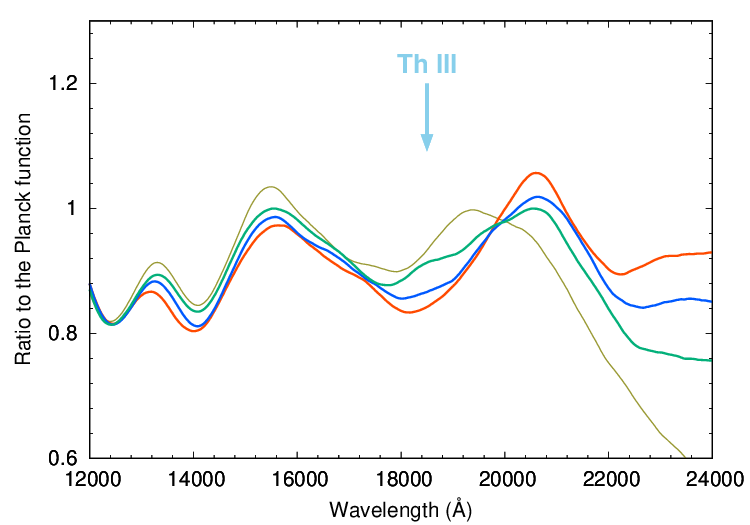} \\
    \end{tabular}
\caption{
  \label{fig:spec-fid}
  Left: synthetic spectra (smoothed) with the fiducial $X$(lan) and various $X$($Z\ge82$) at $t=1.5$ (top) and 2.5 (bottom) days.
  The dark-orange, blue, and green curves show the spectra for the ratio of actinides to lanthanides being 3, 1, and 1/3 of the fiducial ratio, respectively, and the dark-yellow curve shows the spectrum without Th III lines. 
  The gray curve shows the observed spectrum of AT2017gfo at $t=1.5$ and 2.5 days after the merger for comparison purposes \citep{Pian2017, Smartt2017}.\
  The dashed black line shows the Planck function ($T=5300$ K, arbitrary normalized).
  Right: ratios of the synthetic spectra (smoothed) to the Planck function shown in the left panel.
}
\end{center}
\end{figure*}
%===========================

We also evaluate the Soboelv optical depths of bound-bound transitions for the temperature of $T=4000$ K.
This is a typical value in the line-forming region for the L model at $\sim$3 days after the merger (see Figure \ref{fig:temp} and Section \ref{sec:spec}).
The results are shown in the bottom panel of Figure \ref{fig:tau}.
We find that the Th III lines become relatively weak and almost hidden by the other lines at $T=4000$ K.
This is in contrast to the fact that the Ce III lines remain strong compared to the other lines, although their Sobolev optical depths also become smaller.

The reason why the Th III lines rapidly become weaker than the Ce III lines at lower temperature can be understood from their temperature dependences for ionization.
The ionization potentials for Ce are 10.96 eV (II to III) and 20.2 eV (III to IV), and those for Th are 12.10 eV (II to III) and 18.32 eV (III to IV).
Figure \ref{fig:ion} shows the ionization fractions of Ce and Th for the density of $\rho=10^{-14}$ g~cm$^{-3}$ as a function of temperature under LTE.
We find that the temperature range that Th III exists is narrower than that for Ce III.
For $T=4000$ K, most Th ions become singly ionized (II), while $\sim$30\% of Ce III remains.
Thus, the decreased fraction of doubly ionized state results in the weak Th III lines at $T=4000$ K.

This implies that, even if the Th III features appear with the appropriate photospheric temperature at a certain epoch, they would rapidly disappear as the ejecta temperature decreases with time.
We discuss the temperature dependence for the Th III lines on the spectra in Section \ref{sec:result} in more depth.

%===========================
% Figure 
%===========================
\begin{figure*}[th]
  \begin{center}
    \begin{tabular}{cc}
    \includegraphics[width=0.48\linewidth]{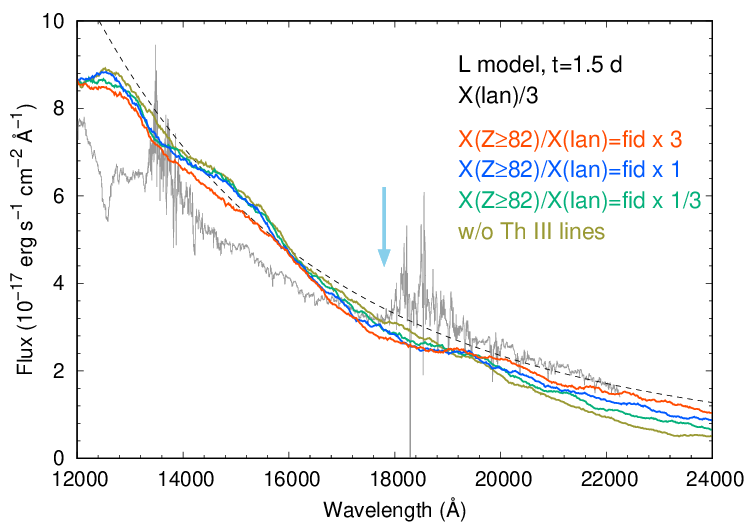} &
    \includegraphics[width=0.48\linewidth]{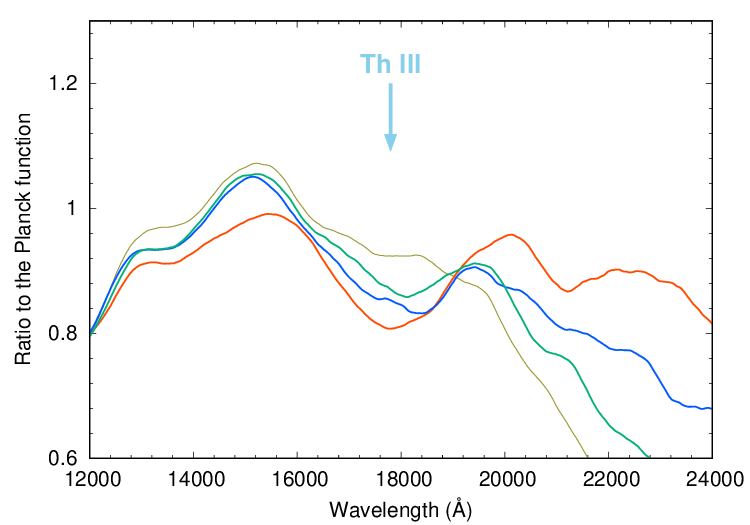} \\
    \includegraphics[width=0.48\linewidth]{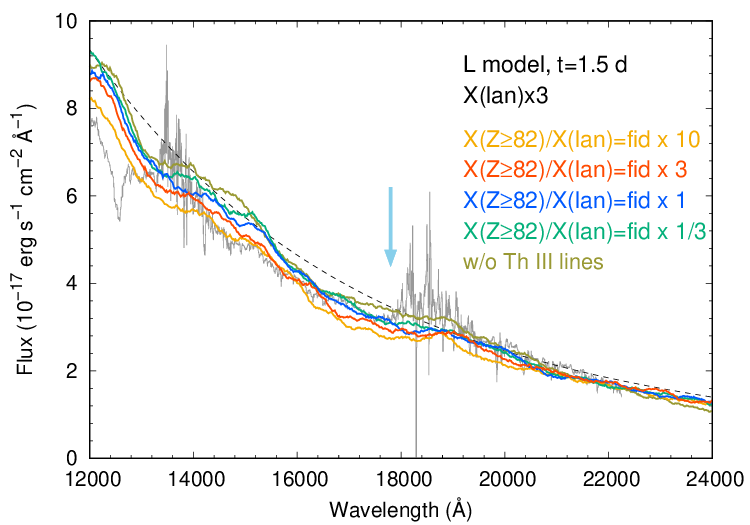} &
    \includegraphics[width=0.48\linewidth]{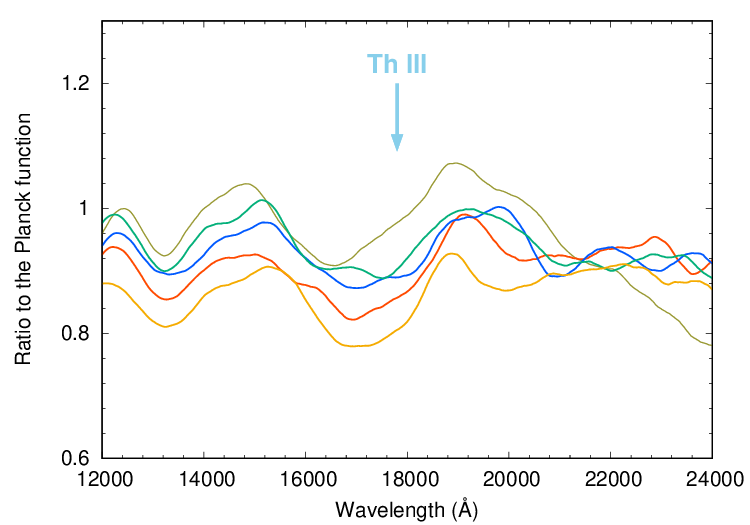} \\
    \end{tabular}
\caption{
  \label{fig:spec-dlan}
  Same as the top panels of Figure \ref{fig:spec-fid}, but with the $X$(lan) reduced (top) or increased (bottom) by a factor of 3 from the original value in the L model.
  The temperature of the Planck function is $T=6000$ and 4500 K for the top and bottom panels, respectively.
}
\end{center}
\end{figure*}
%===========================

\section{Synthetic Spectra}
\label{sec:spec}
\subsection{Fiducial cases}
\label{sec:method}
In this section, we calculate synthetic spectra of kilonovae to investigate the absorption features caused by Th III. 
We use a wavelength-dependent radiative transfer simulation code \citep{TH2013, Tanaka2014, Tanaka2017, Kawaguchi2018, Kawaguchi2020}.
Photon transfer is calculated by the Monte Carlo method.
For the atomic data to compute the opacity for bound-bound transitions, we use the line list updated in Section \ref{sec:linelist}.
For the detailed treatment to compute the bound-bound opacity, we refer the readers to \citet{Domoto2022}.
We assume the ejecta structure with a single power law ($\rho\propto r^{-3}$) for the velocity range of ejecta $v=0.05$--$0.3\ c$ \citep[e.g.,][]{Metzger2010}.
The total ejecta mass is set to be $\Mej = 0.03\Msun$, which is suggested to explain the observed luminosity of AT2017gfo \citep[e.g.,][]{Tanaka2017, Kawaguchi2018}.
 
Here, we use the L model described in Section \ref{sec:strength} as the fiducial model for the abundance distribution of the ejecta.
For comparison purposes, we also use the Solar (S) model adopted in \citet{Domoto2021}, which fits the $r$-process residuals for $A\ge88$ (the green curves in Figure \ref{fig:abun}).
Note that we set the velocity range of ejecta $v=0.05$--$0.4\ c$ for the S model, to avoid that the photosphere at the phase of interest locates at the edge of the ejecta due to the high opacity.
This results in the subtle differences in density structures between models but do not affect the discussion.

The heating rate of radioactive nuclei as a function of time is consistently taken from each abundance model.
The thermalization efficiencies of $\gamma$-rays and radioactive particles follow the analytic formula given by \citet{Barnes2016}.
In the radiative transfer code, the temperature in each cell is determined by the photon flux assuming that the wavelength-integrated photon intensity $\langle J \rangle = \int J_\nu d \nu$ follows the Stefan-Boltzmann law, \ie $\sigma T_R^4=\pi\langle J \rangle$ \citep{Lucy2003, TH2013}.
The kinetic temperature of electrons $T_e$ is assumed to be the same as the radiation temperature $T_R$, \ie $T=T_e =T_R$ under LTE.
The resulting temperature structures of the ejecta for the L and S models are shown in Figure \ref{fig:temp}.

Figure \ref{fig:tau-spec} shows the synthetic spectra at $t=1.5$ days after the merger for the L (left) and S (right) models.
The dark-blue and black curves are the original and smoothed results, respectively.
To find the contributions of relevant elements to the spectra, we also plot the Sobolev optical depths in the photospheric region at $v\sim0.13\ c$ for the L model and at $v\sim0.37\ c$ for the S model, respectively.
The wavelengths of the lines are blueshifted according to the photospheric velocities.
Note that we plot only spectroscopically accurate lines for the L model, while we plot all the lines whose wavelengths are not necessarily accurate for the S model.
To clarify the effects of Th III lines, we compare the results with the cases excluding Th III lines (dark-yellow curve).

For the L model (the left panel), we find that a broad absorption feature appears around $\lambda\sim 18000$ {\AA} (light-blue arrow).
The clear absorption features caused by Ca II, Sr II, La III, and Ce III (blue, red, pink, and orange arrows, respectively) are also seen in the spectrum, which are unaffected by the Th III lines.
This is consistent with the results in \citet{Domoto2022}, although the updated line list of Th III is adopted here.

On the other hand, for the S model (the right panel), we find that the Th III lines decrease overall fluxes by $\sim$10\% but do not produce any additional features.
The spectrum is redder and dimmer than that in the L model. 
It should be noted that the wiggly features seen in the spectra are caused by the lines of heavy elements whose wavelengths are not experimentally calibrated.

These differences between the S and L models reflects the larger mass fraction of lanthanides for the former, which results in a higher total opacity.
We confirmed that higher opacity in the S model shifts the photosphere outward compared to that in the L model.
As a result, the photospheric temperature in the S model is lower and Th III lines are weak (Figure \ref{fig:tau}).
Thus, in our models, larger mass fraction of Th is not necessarily an advantage to produce the Th III features (see also Section \ref{sec:discussion}).

To see the time evolution of the spectra, we show the synthetic spectra for the L model at $t=1.5$ and 2.5 days by blue lines in the left panels of Figure \ref{fig:spec-fid}, focusing on the NIR region.
The spectra without Th III lines are also shown by dark-yellow lines.
To see the effects of Th III lines more clearly, the right panels of Figure \ref{fig:spec-fid} show the ratios of synthetic spectra to the Planck function.
It should be cautioned that we do not intend to present the Planck function as the continuum but just as comparison purposes for each spectrum.
We plot only the smoothed spectra for visualization.

It can be seen that the Th III feature at $\lambda\sim18000$ {\AA} becomes weaker (\ie difference between the blue and dark-yellow lines becomes smaller) as time goes.
Although it may still be recognizable at $t=2.5$ days, the feature completely disappears at $t=3.5$ days (not shown in the figure).
This is because of the temperature dependence of the Th III lines as described in Section \ref{sec:temp}.
In our calculations, the temperature of the ejecta at $v>0.1\ c$ becomes below $\sim$4000 K at $\sim$3 days, at which Th III disappears (Figure \ref{fig:ion} and \ref{fig:temp}).
This suggests that, to detect the Th III features in the photospheric spectra, we need early-time observations during the epochs when the ejecta temperature is high enough (see Section \ref{sec:discussion} for more discussion).

\subsection{Conditions for producing Th features}
\label{sec:result}
To explore the conditions in which the Th absorption feature is observable, we vary the mass fractions of lanthanides ($Z=57$--71) and those of elements with $Z\ge82$ (hereafter denoted as $X$(lan) and $X$($Z\ge82$), respectively) from the original values in the L model.
Although we can in principle assume extreme cases with, \eg a much larger amount of actinides, we avoid such cases;
we impose the mass fractions so that the ratio of actinides to lanthanides keeps by a factor of up to 3 to the fiducial ratio.
This is because some metal-poor stars in the Galactic halo, called ``actinide-boost'' stars, exhibit enhanced Th abundances whose ratio to lanthanides are by a factor of up to 3 with respect to the solar ratio \citep{star, Holmbeck2018, Placco2023}.
Note that, as can be seen in the left panel of Figure \ref{fig:abun}, the ratio of Th with respect to lanthanides (at 1 yr) for the L model (the same as that for the S model) is about a factor of 2 higher than that for the solar $r$-residuals.
This indicates that the present ratio for the L model with the half-life of $^{232}$Th (14.05 Gyr) is similar to the solar value, assuming the merger event about, \eg 13 Gyr ago (comparable to the ages of actinide-boost stars).
Thus, the enhancement of $X$($Z\ge82$) up to a factor of 3 is consistent with the range observed in actinide-boost stars.

For all the cases with varying mass fractions, the heating rate from radioactive decay as a function of time is taken from the L model.
As our abundance model is dominated by relatively light $r$-process elements, the heating rate is also dominated by the light nuclei.
Thus, the changes in mass fractions for such heavy elements do not affect the heating rate during the relevant period of time (a few days after the merger).

We note that varying $X$($Z\ge82$) works effectively the same as varying only the mass fraction of Th as our line list does not include the atomic data for the elements with $Z=89$--100 except for Ac III and Th III (Section \ref{sec:linelist}).
The elements with $Z=82$--88 do not strongly contribute to the total opacity either, as expected from their atomic properties \citep{Tanaka2020, Domoto2022}.
\citet{Fontes2023} suggested that other actinides ($Z>90$), especially for Pa ($Z=91$), may impact to the opacity at the NIR wavelengths.
But this depends on the ejecta conditions (\eg density, temperature, and epoch) as well as abundances.
Thus, in the present simulation, those heaviest elements except for Th just work as little-opacity sources in the ejecta and expected not to seriously affect the Th III features.
Nevertheless, the effects of actinides should be investigated in the future.

The left top panel of Figure \ref{fig:spec-fid} and left panels of Figure \ref{fig:spec-dlan} show the synthetic spectra at $t=1.5$ days for the cases with variations of $X$(lan) and $X$($Z\ge82$).
We compare the spectra for a fixed $X$(lan) along with variations of $X$($Z\ge82$) in each panel.
The dark-orange, blue, and green curves show the cases for the ratio of actinides to lanthanides being 3, 1, and 1/3 of the fiducial ratio, respectively.
The dark-yellow curve shows the case without Th III lines.
The right top panel of Figure \ref{fig:spec-fid} and right panels of Figure \ref{fig:spec-dlan} show the ratios of synthetic spectra at $t=1.5$ days to the Planck function in each panel.

For the cases that $X$(lan) are equal to or less than the fiducial value (Figure \ref{fig:spec-fid} and the top panels of Figure \ref{fig:spec-dlan}), we find the absorption features caused by Th III at $\lambda\sim18000$ {\AA} (see light-blue arrows).
The absorption features become stronger as the mass fraction of Th increases.
In fact, the features caused by Th III are clearer for the smaller $X$(lan) and larger ratio of actinides to lanthanides.
This is because there are relatively strong lines of Ce III around $\lambda\sim20000$ {\AA} at rest wavelengths (see Figure \ref{fig:tau}).
For the fiducial case, the Sobolev optical depths of some of the Ce III lines around $\lambda\sim20000$ {\AA} are larger than 1, which is comparable to those of the Th III lines at similar wavelengths.
Thus, when $X$(lan) is smaller, the absorption features by Th III become more pronounced as the Ce III lines become weaker than the Th III lines.
Although we only varied the abundances here, we confirmed that reducing the mass (and therefore the photospheric velocity) does not largely affect the Th III feature.

It is also seen that the flux at $\lambda\gtrsim 20000$ {\AA} is enhanced when Th III lines are included.
We note that this is likely affected by the assumption in the simulations.
For the absorbed photon packet in the radiative transfer simulation, the comoving wavelength is assigned by sampling the emissivity by the Kirchhoff's law, $j_{\lambda}=\alpha_{\lambda}B_{\lambda}(T)$, where $j_{\lambda}$, $\alpha_{\lambda}$, and $B_{\lambda}(T)$ are the emissivity, the absorption coefficient, and the Planck function, respectively.
This assumption approximately expresses line fluorescence when there are sufficient number of lines \citep[e.g.,][]{Kasen2006}.
However, for the present cases, the emissivity in the NIR wavelengths tends to depend on the relatively small number of strong transitions mainly by Ce III and Th III.
Thus, the amount of Th directly affects the emissivity and resulting fluxes at $\lambda \gtrsim 20000$ {\AA}, and its effect is larger for the smaller $X$(lan).

For the case that $X$(lan) is larger than the fiducial value (the bottom panels of Figure \ref{fig:spec-dlan}), while the increased mass fractions of actinides make the overall fluxes smaller, the effects of Th III lines to absorption features are not clear.
For example, when the ratio of actinides to lanthanides is 10 times larger than the fiducial ratio (orange curves in the bottom panels of Figure \ref{fig:spec-dlan}), we barely find the center of the decreased flux ratio being slightly shifted toward $\lambda\sim18000$ {\AA}.
In fact, even if $X$(lan) increases more, the flux decreases overall regardless of wavelength and no clear absorption can be seen.
This is because a larger $X$(lan) results in a higher total opacity as described in the case of the S model.
This makes the photospheric temperature lower and the absorption features of Th III suppressed.

%===========================
% Figure 
%===========================
\begin{figure}[th]
  \begin{center}
    \includegraphics[width=\linewidth]{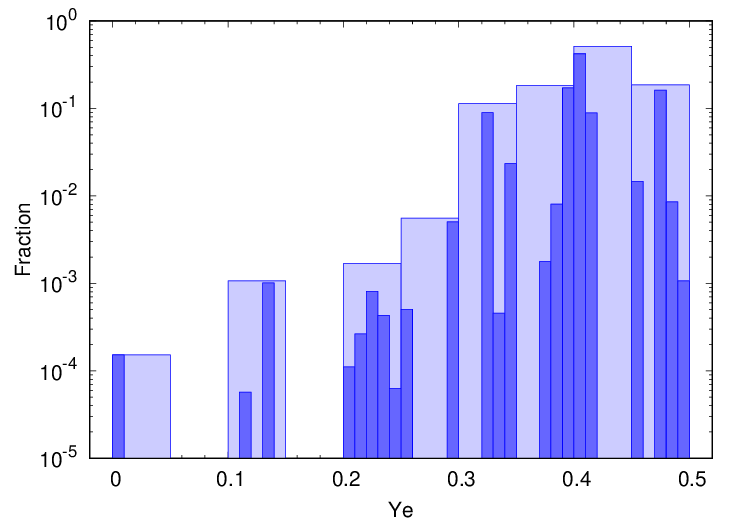}
\caption{
  \label{fig:yehist}
  Histogram of electron fraction $\Ye$ for the L model.
  Denser colors show the histograms with an original interval ($\Delta\Ye=0.01$), while lighter colors show those with a grouped interval ($\Delta\Ye=0.05$).
}
\end{center}
\end{figure}
%===========================

\section{Discussions}
\label{sec:discussion}
Our results suggest that the Th III features appear in the kilonova photospheric spectra if $X$(lan) is smaller than $\sim6\times10^{-4}$ (the fiducial value in the L model) and the ratio of actinides to lanthanides is similar to or larger than the solar $r$-process value in the line-forming region.
As shown in Figure \ref{fig:yehist}, the L model mainly consists of high-$\Ye$ ($>0.3$) components that are responsible for relatively light $r$-process elements, and a small fraction of low-$\Ye$ components that are responsible for heavier elements.
This demonstrates that we may be able to detect the Th III features in the spectra if the bulk of high-$\Ye$ ejecta has such a mixture of low-$\Ye$ components above the photosphere.

Excess of the actinide-to-lanthanide ratio, up to about a factor of 3 with respect to the solar $r$-process value, has been observed in actinide-boost stars, a sub-class (about 1/3) of $r$-process-enhanced stars in the Galactic halo \citep{star, Holmbeck2018, Placco2023}.
Recent nucleosynthesis studies suggest that the mergers of black hole-neutron star binaries \citep{Wanajo2022} or unequal-mass neutron stars \citep{Fujibayashi2023} result in high actinide-to-lanthanide ratios because of the very low $\Ye$ ($\sim$0.04--0.06) in the dynamical ejecta.
The kilonovae associated with the gravitational waves from such merger events may be promising candidates for Th detection in the future.

The main uncertainty in the present results arises from the transition probabilities of the Th III lines.
As the experimental $gf$-values for the NIR lines are not available, we estimated them by using the relative intensities measured by experiments \citep{Engleman2003}.
However, the estimate depends on the temperature adopted using the measured intensities (Equation \ref{eq:I}).
In addition, there is a scatter between measured and calculated values by roughly a factor of 3 (Figure \ref{fig:gf}), as mentioned in Section \ref{sec:linelist}.
The required mass fraction of Th to detect the Th III features roughly scales by the $gf$-values;
if the $gf$-values were larger by a factor of 3, the required mass fraction of Th would become smaller by a factor of 3, and vice versa.
To determine the exact values of the $gf$-values for the Th III lines, more experimental measurements are necessary in the NIR region.

Furthermore, the $gf$-values of Ce III lines are also the source of uncertainty in the Th features.
As discussed in Section \ref{sec:result}, the strength of the Th features ($\lambda\sim 18000$ {\AA}) is affected by the Ce III lines around 20000 {\AA} at rest wavelength.
For the Ce III lines in our line list \citep{Domoto2022}, although the wavelengths and energy levels are experimentally accurate, the $gf$-values are adopted from a theoretical calculation \citep{Tanaka2020} due to the lack of experiments, expect for the strongest lines at $\lambda\sim 16000$ {\AA}\citep{Domoto2023, Gaigalas2024}.
Thus, more experimental measurements for the Ce III lines are also necessary to determine the exact values of the $gf$-values.

It should be noted that the condition to detect the Th III features suggested here is not necessarily the unique solution.
Our models prefer the abundance model dominated by relatively light $r$-process elements for the Th III features to appear in the spectra.
This is because the photospheric temperature lies in an appropriate range for the presence of Th III in the L model but goes outside of the range in the S model.
However, the photospheric temperature in this study depends on the opacity under the assumption of our simplistic homogeneous distribution of the elements.
If the condition of the photospheric temperature $T=5000$--7000 K with a larger mass fractions of Th (and lanthanides) were realized in more realistic ejecta profiles, the Th III features would become even stronger than those presented here \citep[see, e.g.,][for the density and $\Ye$ structure of the ejecta from self-consistent merger simulations]{Kawaguchi2023, Kawaguchi2024}.
This possibility cannot be tested in our framework, because the opacity and temperature are coupled based on the assumed abundance distribution in the simulations.

Another caveat is that our calculations assume LTE.
According to the temperature dependence of the Th III lines under LTE, we need early-time observation to detect the Th III features.
However, the assumption of LTE may not be valid in the low-density regions of ejecta.
Non-LTE effects may modify the ionization faction and temperature structure in the ejecta with time, \ie with decreasing density, which leads overionization \citep{Hotokezaka2021, Pognan2022b}.
For example, it has been shown that $T_e$ can stay around a few 1000~K, which keeps about a few 10 \% of heavy elements in doubly ionized states even at 20 days \citep{Hotokezaka2021, Pognan2023}.
This may allow us to keep finding the Th III features for longer duration than that presented here, \eg even after $\sim$3 days after the merger.

In this paper, we do not attempt to identify Th III in the spectra of AT2017gfo.
Although we showed the spectrum of AT2017gfo for comparison purposes in the left panels of Figure \ref{fig:spec-fid} and \ref{fig:spec-dlan}, it is difficult to judge the existence of the feature around $\lambda\sim18000$ {\AA} due to the strong telluric absorption.
To observationally confirm the absorption features by Th III in the future, observations from the space or high-altitude sites with no or little Earth's atmosphere are necessary.
For a GW170817-like kilonova associated with a detection of GW at $\lesssim200$ Mpc (a design detection limit of GWs from NS mergers for LIGO/Virgo), observations with JWST/NIRSpec \citep{nirspec1, nirspec2} or upcoming telescope at a high altitude such as TAO \citep{tao} will give sufficient signal to noise ratio to find the Th III features in the photospheric spectra.

\section{Conclusions}
\label{sec:conclusion}
We have explored the signatures of the elements beyond the third $r$-process peak in the photospheric spectra of kilonovae.
We constructed the line list of Ra II, Ac III, and Th III with an anticipation that their atomic properties are analogous to those of Ca II/Sr II, La III, and Ce III, respectively, which tend to produce strong absorption features in kilonova spectra.
By systematically calculating the strength of bound-bound transitions using this line list, we have found that
Th III is the most promising species that shows comparably strong lines in the NIR wavelengths to, \eg Ce III lines.
This is because Th III has the dense low-lying energy levels and the lines with relatively large $gf$-values, which make the Sobolev optical depths of the lines relatively large.
Although the Ra II and Ac III lines might stand out from other lines in the optical wavelengths in cases where the light $r$-process elements are absent, we need experimentally calibrated atomic data for more lanthanides to fully verify this possibility.

We performed radiative transfer simulations to obtain synthetic spectra of kilonovae.
We have shown that the Th III lines produce broad absorption features at $\lambda\sim18000$ {\AA} in the spectra since $\sim$1 days after the merger if $X$(lan)$\lesssim 6\times10^{-4}$ and the ratio of actinides to lanthanides larger than the solar $r$-process value in the line-forming region.
This suggests that we may be able to detect such signatures of Th if the bulk of high-$\Ye$ ejecta has a mixture of a small fraction of low-$\Ye$ components above the photosphere.
Such conditions may be realized in the mergers of black hole-neutron star binaries or unequal-mass neutron stars.
If we are able to detect the signatures caused by Th, in turn it becomes the unambiguous evidence that the third $r$-process peak and heavier elements are synthesized in the ejecta of NS mergers.
To detect the Th absorption features, the observations from the space or high-altitude sites are important as the wavelength region of the Th features is overlapped with that affected by the strong telluric absorption.

\begin{acknowledgments}
We thank C.~J. Fontes for helping us to collect the references on the experimental data of Th lines.
Numerical simulations presented in this paper were carried out on Cray XC50 at Center for Computational Astrophysics, National Astronomical Observatory of Japan.
N.D. acknowledges support from Graduate Program on Physics for the Universe (GP-PU) at Tohoku University.
This work was supported by JST FOREST Program (Grant Number JPMJFR212Y, JPMJFR2136), NIFS Collaborative Research Program (NIFS22KIIF005), the Grant-in-Aid for JSPS Fellows (22KJ0317), and the Grant-in-Aid for Scientific Research from JSPS (20H00158, 21H04997, 23H00127, 23H04891, 23H04894, 23H05432).
\end{acknowledgments}

\appendix
\section{Updated transition probabilities of Th III lines}
\label{sec:app}
We summarize the $gf$-values of Th III lines estimated by using the measured line intensities in Table \ref{tab:gf}.
The listed lines are overlapped with those in Table 5 of \citet{Domoto2022} but with the updated $gf$-values (see Section \ref{sec:linelist}).

%===========================
% Table 
%===========================
\startlongtable
\begin{deluxetable*}{lccccccc}%[pht]
\tablewidth{0pt}
\tablecaption{Summary of lines for Th III whose $gf$-values are estimated from the measured line intensities.}
\label{tab:gf}
\tablehead{
   &  $\lambda_{\rm vac}$$^a$  &  $\lambda_{\rm air}$$^b$  & Lower level & $E_{\rm lower}$$^c$ & Upper level & $E_{\rm upper}$$^d$ & log $gf^e$ \\
   &  ({\AA})                                &  ({\AA})                               &                    &  (${\rm cm^{-1}}$)  &                         &   (${\rm cm^{-1}}$)     & 
}
\startdata
Th III  &  10046.6355  &  10043.8821  & 5$f$6$d$ ${\rm ^3H_4^o}$ &      0.00000    & 6$d$7$s$ ${\rm ^3D_3}$    &    9953.58098  & $-$2.192 \\ 
          &  10257.0203  &  10254.2099  & 5$f$6$d$                             &   6288.2212   & 6$d$7$s$ ${\rm ^1D_2}$    &  16037.64127  & $-$1.441 \\ 
          &  10260.9778  &  10258.1662  & 5$f$6$d$ ${\rm ^3G_4^o}$ &   8141.74929 & 5$f^2$ ${\rm ^3H_5}$         &  17887.40925  &     0.606 \\ 
          &  10532.8696  &  10529.9844  & 5$f$6$d$ ${\rm ^3G_5^o}$ &  11276.80704 & 5$f^2$ ${\rm ^3H_6}$         &  20770.89593  &     0.877 \\ 
          &  10581.3571  &  10578.4587  & 5$f$6$d$ ${\rm ^3H_6^o}$ &    8436.8256   & 5$f^2$ ${\rm ^3H_5}$         &  17887.40925  & $-$1.367 \\ 
          &  10706.5771  &  10703.6448  & 5$f$6$d$ ${\rm ^1H_5^o}$ &  19009.90952 & 5$f^2$ ${\rm ^3I_6}$          &  28349.96275  &     1.153 \\ 
          &  11216.3023  &  11213.2317  & 6$d^2$     ${\rm ^3F_4}$     &   6537.81726  & 5$f$6$d$ ${\rm ^1F_3^o}$ &  15453.41152  & $-$1.118 \\ 
          &  11227.3116  &  11224.2381  & 5$f$6$d$                              &   8980.55749  & 5$f^2$ ${\rm ^3H_5}$         &  17887.40925  &     0.140 \\ 
          &  11428.8103  &  11425.6821  & 7$s^2$      ${\rm ^1S_0}$    &  11961.1316    & 5$f$6$d$ ${\rm ^1P_1^o}$ &  20710.9487   & $-$0.916 \\ 
          &  11516.6200  &  11513.4679  & 5$f$6$d$ ${\rm ^3D_2^o}$  &  10180.76952 & 5$f^2$ ${\rm ^3F_2}$         &  18863.86913  & $-$1.192 \\ 
          &  11720.8814  &  11717.6740  & 6$d^2$                                 &    4676.43214  & 5$f$6$d$ ${\rm ^3P_2^o}$ &  13208.2137  & $-$1.912 \\ 
          &  11810.5436  &  11807.3118  & 6$d^2$       ${\rm ^1G_4}$   &  10542.89973  & 5$f$6$d$ ${\rm ^1H_5^o}$ &  19009.90952  &     0.260 \\ 
          &  12081.2271  &  12077.9218  & 6$d$7$s$ ${\rm ^3D_2}$    &    7176.10661  &  5$f$6$d$ ${\rm ^1F_3^o}$ &  15453.41152  & $-$1.481 \\ 
          &  12320.5004  &  12317.1302  & 5$f$6$d$ ${\rm ^3D_1^o}$  &   7921.08782  &  6$d$7$s$ ${\rm ^1D_2}$    &  16037.64127  & $-$0.629 \\ 
          &  12726.1743  &  12722.6939  & 6$d^2$     ${\rm ^3F_2}$     &        63.26679  &  5$f$6$d$ ${\rm ^3D_1^o}$ &   7921.08782  & $-$1.374 \\ 
          &  12918.7449  &  12915.2122  & 5$f$6$d$ ${\rm ^3P_1^o}$  &  11123.1791    & 5$f^2$ ${\rm ^3F_2}$          &  18863.86913  & $-$1.117 \\ 
          &  13075.4606  &  13071.8853  & 5$f$7$s$                              &   7500.60523  & 5$f^2$ ${\rm ^3H_4}$          &  15148.51938  & $-$0.782 \\ 
          &  13102.2536  &  13098.6710  & 5$f$6$d$ ${\rm ^3P_2^o}$  &  13208.2137   & 5$f^2$ ${\rm ^3F_3}$          &  20840.48853  & $-$1.095 \\ 
          &  13445.6702  &  13441.9944  & 6$d^2$       ${\rm ^3F_2}$   &        63.26579  & 5$f$7$s$                             &   7500.60523  & $-$2.294 \\ 
          &  13465.3195  &  13461.6384  & 5$f$7$s$                              &    2527.09550  & 6$d$7$s$ ${\rm ^3D_3}$    &   9953.58098  & $-$1.298 \\ 
          &  13577.6105  &  13573.8999  & 5$f$6$d$ ${\rm ^3F_2^o}$  &      510.75823  & 6$d^2$       ${\rm ^3P_1}$   &   7875.8244    & $-$2.783 \\ 
          &  13596.9359  &  13593.2199  & 5$f$6$d$                             &    3188.30107  & 6$d^2$       ${\rm ^1G_4}$   &  10542.89873  & $-$2.180 \\ 
          &  14271.9111  &  14268.0108  & 5$f$6$d$ ${\rm ^3G_4^o}$  &    8141.74929  & 5$f^2$       ${\rm ^3H_4}$    &  15148.51938  & $-$1.264 \\ 
          &  14363.1448  &  14359.2196  & 5$f$6$d$ ${\rm ^1H_5^o}$  &  19009.90952  & 5$f^2$       ${\rm ^1G_4}$    &  25972.17312  & $-$0.063 \\ 
          &  14766.5150  &  14762.4802  & 5$f$7$s$ ${\rm ^3F_2^o}$   &    3181.5024   & 6$d$7$s$ ${\rm ^3D_3}$     &    9953.58098  & $-$2.494 \\ 
          &  14781.3544  &  14777.3156  & 5$f$6$d$                              &    3188.30107  & 6$d$7$s$ ${\rm ^3D_3}$    &    9953.58098  & $-$1.406 \\ 
          &  14958.5620  &  14954.4751  & 6$d^2$     ${\rm ^3F_3}$      &   4056.01548  & 5$f$6$d$ ${\rm ^3D_3^o}$  &  10741.15004  & $-$1.863 \\ 
          &  15002.9667  &  14998.8677  & 5$f$6$d$ ${\rm ^3F_2^o}$   &     510.75823  & 6$d$7$s$ ${\rm ^3D_2}$     &   7176.10661  & $-$1.468 \\ 
          &  15127.2149  &  15123.0820  & 5$f$6$d$ ${\rm ^3G_5^o}$   &  11276.80704  & 5$f^2$ ${\rm ^3H_5}$         &  17887.40925  & $-$1.056 \\ 
          &  15295.6248  &  15291.4462  & 5$f$6$d$ ${\rm ^3H_4^o}$   &         0.00000  & 6$d^2$ ${\rm ^3F_4}$         &    6537.81726  & $-$1.660 \\ 
          &  15511.6993  &  15507.4619  & 6$d^2$                                  &    4676.43214  & 5$f$6$d$ ${\rm ^3P_1^o}$ &  11123.1791    & $-$1.650 \\ 
          &  16064.3737  &  16059.9870  & 6$d^2$     ${\rm ^3F_2}$      &        63.26579  & 5$f$6$d$                             &   6288.2212    & $-$2.912 \\ 
          &  16212.8109  &  16208.3829  & 5$f$6$d$                              &    8980.55749  & 5$f^2$ ${\rm ^3H_4}$          &  15148.51938  & $-$1.981 \\ 
          &  16327.1959  &  16322.7367  & 6$d^2$     ${\rm ^3F_3}$      &    4056.01548  & 5$f$6$d$ ${\rm ^3D_2^o}$  &  10180.76592  & $-$1.451 \\ 
          &  16488.8131  &  16484.3100  & 6$d^2$                                  &   4676.43214  & 5$f$6$d$ ${\rm ^3D_3^o}$  &  10741.15004  & $-$0.608 \\ 
          &  16577.9550  &  16573.4277  & 6$d$7$s$ ${\rm ^3D_2}$      &   7176.10661  & 5$f$6$d$ ${\rm ^3P_2^o}$  &  13208.2137    & $-$1.647 \\ 
          &  17073.9505  &  17069.2883  & 5$f$6$d$ ${\rm ^3D_2^o}$   &  10180.76592  & 6$d$7$s$ ${\rm ^1D_2}$    &  16037.64127  & $-$1.647 \\ 
          &  17494.5296  &  17489.7530  & 5$f$6$d$                               &    4826.82620  & 6$d^2$       ${\rm ^1G_4}$  &  10542.89873  & $-$0.961 \\ 
          &  17517.0190  &  17512.2362  & 6$d$7$s$ ${\rm ^3D_1}$      &    5523.8809   & 5$f$6$d$ ${\rm ^3P_0^o}$   &  11232.6148    & $-$1.675 \\ 
          &  17814.4804  &  17809.6167  & 5$f$6$d$                               &    4826.82620 & 6$d^2$       ${\rm ^3P_2}$    &  10440.2372   & $-$2.535 \\ 
          &  17859.3810  &  17854.5071  & 6$d$7$s$ ${\rm ^3D_1}$      &   5523.8809    & 5$f$6$d$ ${\rm ^3P_1^o}$    &  11123.1791   & $-$1.699 \\ 
          &  18182.3784  &  18177.4147  & 6$d$7$s$ ${\rm ^3D_3}$      &   9953.58089  & 5$f$6$d$ ${\rm ^1F_3^o}$    &  15453.41152  & $-$2.694 \\ 
          &  18240.3369  &  18235.3574  & 5$f$6$d$ ${\rm ^3G_3^o}$   &   5060.54386  &  6$d^2$       ${\rm ^1G_4}$   &  10542.89973  & $-$1.712 \\ 
          &  18588.4201  &  18583.3459  & 5$f$6$d$ ${\rm ^3G_3^o}$   &   5060.54386  &  6$d^2$       ${\rm ^3P_2}$   &  10440.2372    & $-$2.499 \\ 
          &  18880.4240  &  18875.2703  & 5$f$6$d$ ${\rm ^3D_3^o}$   &  10741.15004  & 6$d$7$s$ ${\rm ^1D_2}$     &  16037.64127  & $-$1.647 \\ 
          &  19947.4412  &  19941.9972  & 6$d^2$     ${\rm ^3P_2}$      &  10440.2372   & 5$f$6$d$ ${\rm ^1F_3^o}$   &  15453.41152  & $-$0.500 \\ 
          &  19947.6467  &  19942.2026  & 5$f$6$d$ ${\rm ^3F_2^o}$   &      510.75823  & 6$d$7$s$ ${\rm ^3D_1}$     &   5523.8809    & $-$1.201 \\ 
          &  20010.8977  &  20005.4364  & 6$d^2$     ${\rm ^3F_2}$      &        63.26679  & 5$f$6$d$ ${\rm ^3G_3^o}$   &   5060.54386  & $-$1.296 \\ 
          &  20306.4569  &  20300.9151  & 6$d^2$     ${\rm ^3F_3}$      &    4056.01548  & 5$f$6$d$                               &   8980.55749  & $-$0.629 \\ 
          &  20364.4720  &  20358.9144  & 6$d^2$     ${\rm ^1G_4}$      &  10542.89873 & 5$f$6$d$ ${\rm ^1F_3^o}$   &  15453.41152  & $-$0.596 \\ 
          &  20437.2044  &  20431.6271  & 5$f$6$d$ ${\rm ^3G_3^o}$   &   5060.54386  & 6$d$7$s$ ${\rm ^3D_3}$      &   9953.58098  & $-$1.990 \\ 
          &  20992.7054  &  20986.9770  & 6$d^2$     ${\rm ^3F_2}$       &       63.26679  & 5$f$6$d$                               &   4826.82620  & $-$0.844 \\ 
          &  21101.5437  &  21095.7856  & 6$d^2$     ${\rm ^3F_4}$       &    6537.81726  & 5$f$6$d$ ${\rm ^3G_5^o}$  &  11276.80704  & 0.079 \\ 
          &  21398.121    &  21392.283    & 6$d$7$s$ ${\rm ^1D_2}$      &  16037.64127  & 5$f$6$d$ ${\rm ^1P_1^o}$   &  20710.9487   & $-$0.683 \\ 
          &  21473.5815  &  21467.7221  & 6$d$7$s$ ${\rm ^3D_1}$      &   5523.8809    & 5$f$6$d$ ${\rm ^3D_2^o}$   &  10180.76592  & $-$3.099 \\ 
          &  21509.9507  &  21504.0814  & 5$f$7$s$                               &   2527.09550  & 6$d$7$s$ ${\rm ^3D_2}$      &   7176.10661  & $-$1.132 \\ 
          &  22689.2716  &  22683.0813  & 5$f$6$d$ ${\rm ^3D_3^o}$   &  10741.15004  & 5$f^2$       ${\rm ^3H_4}$    &  15148.51938  & $-$1.492 \\ 
          &  23628.9815  &  23622.5353  & 5$f$7$s$ ${\rm ^3F_4^o}$   &    6310.80773  & 6$d^2$       ${\rm ^1G_4}$   &  10542.89873  & $-$1.487 \\ 
          &  23790.6455  &  23784.1553  & 6$d^2$     ${\rm ^3F_4}$      &    6537.81726  & 5$f$6$d$ ${\rm ^3D_3^o}$  &  10741.15004  & $-$0.956 \\ 
          &  24005.7196  &  23999.1709  & 5$f$6$d$ ${\rm ^3F_2^o}$   &      510.75823  & 6$d^2$                                 &   4676.43214  & $-$1.534 \\ 
          &  24475.4075  &  24468.7309  & 6$d^2$     ${\rm ^3F_3}$       &    4056.01548  & 5$f$6$d$ ${\rm ^3G_4^o}$ &   8141.74929  & $-$0.156 \\ 
          &  25335.2327  &  25328.3219  & 6$d$7$s$ ${\rm ^3D_2}$      &    7176.10661  & 5$f$6$d$ ${\rm ^3P_1^o}$  &  11123.1791  & $-$1.816 \\ 
          &  25897.4091  &  25890.3452  & 5$f^2$ ${\rm ^3H_4}$           &  15148.51938  & 5$f$6$d$ ${\rm ^1H_5^o}$  &  19009.90952  & $-$1.317 \\ 
          &  27282.451    &  27275.010    & 5$f$6$d$                               &    6288.2212    & 6$d$7$s$ ${\rm ^3D_3}$     &   9953.58098  & $-$2.915 \\ 
          &  27451.6126  &  27444.1255  & 5$f$7$s$ ${\rm ^3F_4^o}$   &    6310.80773  & 6$d$7$s$ ${\rm ^3D_3}$      &   9953.58098  & $-$0.763 \\ 
          &  28050.1492  &  28042.4991  & 6$d$7$s$ ${\rm ^3D_2}$      &    7176.10661  & 5$f$6$d$ ${\rm ^3D_3^o}$  &  10741.15004  & $-$1.290 \\ 
          &  28206.6978  &  28199.0051  & 5$f$6$d$ ${\rm ^3F_2^o}$   &      510.75823  & 6$d^2$     ${\rm ^3F_3}$       &   4056.0148   & $-$2.009 \\ 
          &  29790.362    &  29782.238    & 6$d^2$     ${\rm ^3P_1}$       &   7875.8244    & 5$f$6$d$ ${\rm ^3P_0^o}$   &  11232.6148   & $-$2.705 \\ 
          &  29855.0580  &  29846.9163  & 5$f$6$d$                               &   3188.30107  & 6$d^2$     ${\rm ^3F_4}$       &   6537.81726  & $-$1.151 \\ 
          &  30725.433    &  30717.054    & 6$d$7$s$ ${\rm ^3D_3}$      &   9953.58098  & 5$f$6$d$ ${\rm ^3P_2^o}$   &  13208.2137  & $-$1.194 \\ 
          &  30794.295    &  30785.898    & 6$d^2$     ${\rm ^3P_1}$       &   7875.8244    & 5$f$6$d$ ${\rm ^3P_1^o}$   &  11123.1791  & $-$1.409 \\ 
          &  30819.9112  &  30811.5068   & 6$d^2$                                   &   4676.43214  & 5$f$6$d$ ${\rm ^3D_1^o}$   &   7921.08782  & $-$1.623 \\ 
          &  32069.418    &  32060.673    & 6$d^2$     ${\rm ^3F_2}$       &       63.26679  & 5$f$7$s$ ${\rm ^3F_2^o}$    &   3181.5024   & $-$2.526 \\ 
          &  32869.938    &  32860.976    & 5$f$7$s$                               &    7500.60523  & 6$d^2$       ${\rm ^1G_4}$   &  10542.89873  & $-$1.330 \\ 
          &  33281.644    &  33272.569    & 6$d$7$s$ ${\rm ^3D_2}$      &    7176.10661  & 5$f$6$d$ ${\rm ^3D_2^o}$   &  10180.76592 & $-$1.764 \\ 
          &  34017.864    &  34008.588    & 5$f$7$s$                               &    7500.60523  & 6$d^2$     ${\rm ^3P_2}$      &  10440.2372   & $-$1.558 \\ 
          &  35322.815    &  35313.184    & 6$d^2$     ${\rm ^3P_0}$       &   5090.0564    & 5$f$6$d$ ${\rm ^3D_1^o}$   &   7921.08782  & $-$1.411 \\ 
          &  35408.695    &  35398.940    & 6$d^2$                                   &   4676.43214  & 5$f$7$s$                              &   7500.60523  & $-$1.236 \\ 
          &  36127.474    &  36117.624    & 6$d^2$     ${\rm ^3P_2}$       &  10440.2372   & 5$f$6$d$ ${\rm ^3P_2^o}$  &  13208.2137    & $-$2.332 \\ 
          &  40587.237    &  40576.172    & 6$d^2$     ${\rm ^3F_2}$       &        63.26679  & 5$f$7$s$                              &   2527.09550  & $-$1.042 \\ 
          &  40766.811    &  40755.697    & 5$f$7$s$                               &    7500.60523  & 6$d$7$s$ ${\rm ^3D_3}$      &   9953.58098  & $-$2.566 \\ 
          &  40937.632    &  40926.471    & 6$d^2$     ${\rm ^3F_4}$       &   6537.81726  &  5$f$6$d$                              &   8980.55749  & $-$1.965 \\ 
          &  41715.214    &  41703.842    & 6$d$7$s$ ${\rm ^3D_1}$      &    5523.8809   & 5$f$6$d$ ${\rm ^3D_1^o}$   &   7921.08809  & $-$2.281 \\ 
          &  42566.226    &  42554.621    & 5$f$6$d$                               &    4826.82620  & 6$d$7$s$ ${\rm ^3D_2}$      &   7176.10661  & $-$2.341 \\ 
          &  42691.649    &  42680.010    & 5$f$7$s$ ${\rm ^3F_2^o}$    &    3181.5024    & 6$d$7$s$ ${\rm ^3D_1}$      &   5523.8809   & $-$1.443 \\ 
          &  43385.049    &  43373.221    & 6$d^2$     ${\rm ^3P_1}$       &   7875.8244    & 5$f$6$d$ ${\rm ^3D_2^o}$   &  10180.76592  & $-$1.090 \\ 
          &  44349.984    &  44337.894    & 6$d^2$     ${\rm ^3F_3}$       &   4056.01548  & 5$f$7$s$ ${\rm ^3F_4^o}$    &   6310.80773  & $-$1.474 \\ 
          &  46525.983    &  46513.300    & 5$f$7$s$                               &   2527.09550  & 6$d^2$                                   &   4676.43214  & $-$1.510 \\ 
\enddata
\tablecomments{
$^a$ Vacuum transition wavelength. \\
$^b$ Air transition wavelength. \\
$^c$ Lower energy level. \\
$^d$ Upper energy level. \\
$^e$ $gf$-value estimated in Section \ref{sec:linelist}.
}
\end{deluxetable*}
%===========================

\bibliography{references}
\bibliographystyle{aasjournal}

\end{document}